# Title: Mechanics of Interfacial Composite Materials

**Authors:** Anand Bala Subramaniam, Manouk Abkarian, L. Mahadevan & Howard A. Stone

**Affiliation:** *Division of Engineering and Applied Sciences, Harvard University, Cambridge, MA 02138*

**Corresponding Author:**

Howard A. Stone

Vicky Joseph Professor of Engineering and Applied Mathematics

Division of Engineering and Applied Sciences, 308 Pierce Hall
Harvard University
Cambridge, MA  02138
has@deas.harvard.edu
phone: (617) 495-3599;  fax: (617) 495-9837

**Abbreviations:** ICM: Interfacial Composite Material

## Abstract

Recent experiments and simulations have demonstrated that particle-covered interfaces can exist in stable non-spherical shapes as a result of the steric jamming of the interfacially trapped particles, which confers the interface with solid-like properties. We provide an experimental and theoretical characterization of the mechanical properties of these armored objects, with attention given to the two-dimensional granular state of the interface. Small inhomogeneous stresses produce a plastic response while homogeneous stresses produce a weak elastic response. Shear-driven particle-scale rearrangements explain the basic threshold needed to obtain the near-perfect plastic deformation that is observed. Furthermore, the inhomogeneous stress state of the interface is exhibited experimentally by using surfactants to destabilize the particles on the surface. Since the interfacially trapped particles retain their individual characteristics, armored interfaces can be recognized as a kind of composite material with distinct chemical, structural and mechanical properties.

## Introduction

It is well known in the materials world that one can create new materials with unique properties that are different from their constituent ingredients[1]. Such principles can also be used to create new kinds of interfaces with their own distinct properties. Indeed, because interfaces are important for catalysis[2], mass and heat exchange, structural stability, encapsulation, adhesion, chemical and thermal sensing among others, the design of novel interfaces is an active area of research in the materials and colloid sciences. Recently, various experiments and simulations have shown that armored interfaces-fluid/fluid interfaces with rigid particles-are able to support bubbles and drops with non-spherical shapes and other non-minimal surfaces[3-5]. The jamming of the interfacially trapped particles results in solid-like properties that allows the support of these non-minimal shapes[3-6]. As we describe in this paper, these armored interfaces have the attributes of composite materials: the fluid/fluid interface and the individual particles maintain their distinct properties (chemistry, size, shape), while giving rise to collective mechanical properties. We thus describe these systems generally as *interfacial composite materials (*ICMs).

The concept of jamming on interfaces as the origin for the solid-like properties of the ICMs has been explored relatively recently despite the long history of the use of particles as stabilizers for emulsions and foams[7, 8]. We note that non-spherical shapes of particle-covered droplets have been observed occasionally as by-products of emulsification[7, 9], or even due to collection of organic matter on bubbles in the ocean[10], with no further studies on their physical origin or mechanical properties. Particle-covered interfaces have also been used to make chemically or thermally sintered shells of controlled permeability[11, 12]. In addition, particles have been used to stabilize single microbubbles against disproportionation[13]. Finally as an example of composite design, a microfluidic approach has been used to make spherical Janus ICMs, which have regions with distinct particle size and chemistry[6]. For a review of the subject of particles on interfaces as of 2002, see Binks[14]. In particular, there have been many experimental and theoretical studies of armored interfaces that focus on isotropic mechanical properties on planar interfaces. Various studies have reported experiments and/or modeling of the Young's modulus[15, 16], bending rigidity[17] and surface pressure[18, 19] of particulate monolayers.

Here we provide an experimental and theoretical characterization of the mechanical properties of ICMs on closed compact surfaces. We employ an experimental system of micron to millimeter size non-spherical bubbles; the spheroidal geometry of the bubble allows the identification of locally inhomogeneous mechanical behaviors that are not apparent in planar geometries. We find that the ICM responds plastically to relatively small inhomogeneous stresses, while responding rigidly when subject to homogenous stresses. We interpret these experimental observations in terms of shear-driven particle-scale rearrangements, and characterize these systems as two-dimensional granular solids. Finally, we use the closed spheroidal geometry of the bubble to deduce the anisotropic stress distribution of these objects through the destabilization of the ICM by surfactants.

## Experimental Methods

**Particles and fluids:** We made air bubbles and mineral oil (Sigma, research grade) droplets, which were dispersed in water. We used monodisperse polystyrene particles of diameter 1.6 μm, 2.1 μm, 4.0 μm, 4.6 μm, 100 μm, and 500 μm, 1.6 μm diameter silica particles (Bangs Lab), 1.0 μm diameter PMMA particles (Bangs Lab), polydisperse agglomerated gold microparticles with mean diameters ranging from 1.0–3.99 μm (Sigma), and ground ZrO particles (Glenn Mills) with diameters ranging from 200-400 μm.

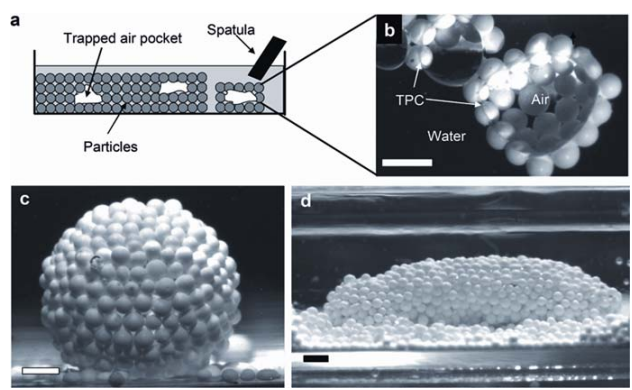

**Figure 1:** 'Air-pocket-trapping' technique for synthesizing bubbles covered with granular particles greater than 10 μm in diameter. (a) A layer of ZrO particles is placed into a petri dish and water is added rapidly to the dish. The water infiltrates the porous layer and traps pockets of air between the particles, which forms three-phase-contact lines on the particles. (b) An experimental image of an armored bubble adhered to the petri dish, which shows the configuration of the particles on the interface. The particles with a contact line are held strongly by the interface and thus gentle sweeping with a spatula removes particles that do not have a contact line. The spatula can then be used to shape the trapped pockets into (c) spherical armored bubbles, or (d) non-spherical armored bubbles. Here, the granular particles used were heavy, and the armor of particles weighs down the air bubbles. Scale bars 400 μm.

**Production of armored bubbles:** Spherical bubbles armored with spherical particles of diameter less than 10 μm were produced through a hydrodynamic flow focusing method[6]. Non-spherical bubbles were produced by forced coalescence[3]. Bubbles covered with hydrophilic particles greater than 10 μm in diameter were prepared through a novel "air pocket trapping" technique. ZrO particles were manually sorted to increase monodispersity and poured into a petri dish to form a thin porous layer. Distilled water (Millipore) was quickly added to the particle layer to ensure entrapment of air between the particles (Fig. 1a). The water penetrates the porous layer until further infiltration is halted when the capillary pressure of the advancing water interface is balanced by the pressure of the trapped air, resulting in pinned three-phase contact lines on the particles. In this way, a monolayer of particles is trapped at the air/water interface, and the untrapped particles can be brushed away with a spatula (Fig. 1a,b). The trapped air pockets can then be shaped with a spatula to produce spherical and non-spherical coated bubbles (Fig. 1c,d).

**Shear flow experiment for verification of bubble rigidity:** The armored bubbles were transferred to a continuous medium of pure glycerol to achieve higher shear stresses during flow. The flow chamber employed was a custom-made quartz parallelepiped with a height of 1.0 mm, width 10 mm and length of 10 mm. Shear rates of up to 4 s⁻¹ were applied by controlling the flow rate with a syringe pump.

**Cryo-SEM imaging of ICM:** A solution containing armored bubbles was introduced onto a copper sample grid and flash frozen in liquid ethane at -180°C. The resultant material containing amorphous ice was shattered with a microblade and the sample was sputter coated with a 10 nm layer of gold.

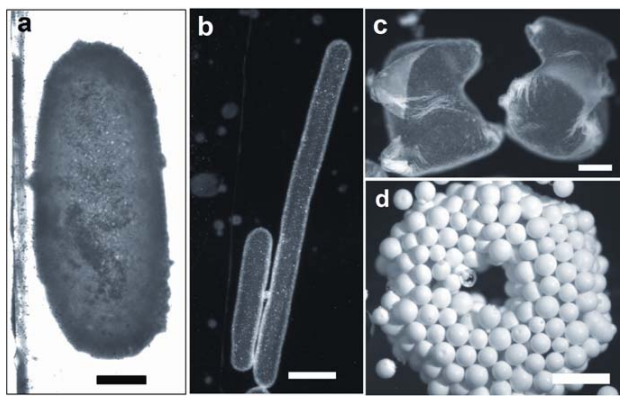

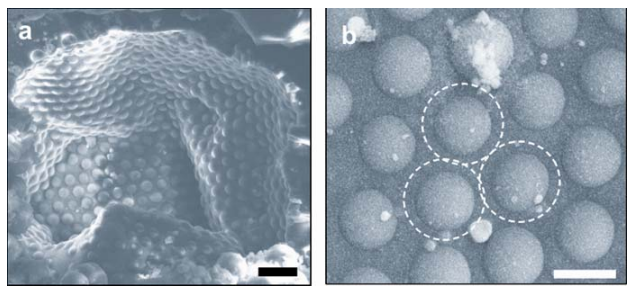

**Figure 2:** Stable support of different anisotropic shapes with various particle sizes and bubble sizes. (a) A stable ellipsoidal bubble obtained by fusing two spherical armored bubbles. Scale bar 100 µm. (b) The mechanical properties of the ICM allows the support of extremely high-aspect ratio gas bubbles such as these millimetre length spherocylinders formed from successive coalescence events. Scale bar 400 µm. (c) A membrane-like solid created with 200 nm fluorescent polystyrene particles by partially evacuating an initially spherical bubble. Scale bar 200 µm. (d) The ability to maintain saddle curvature allows a permanent change of topology of an air bubble by introducing a hole to produce a genus 1 toroid. Particles are ground ZrO with an average diameter of 200 µm. Scale bar 500 µm. Part (a) and (d) appeared in reference (3).

**Figure 3:** Structure of the ICM. (a) Freeze fracture SEM image of a bubble armoured with 4.0 µm diameter polystyrene beads. The rapid freezing preserves the amorphous liquid structure of the air/water interface around the embedded particles. The area viewed is the interior of the bubble, which has been cut open. Scale bar 12 µm. (b) A close-up of the interfacial solid, showing the penetration depth of the particles. We estimate a particle contact angle of 40 degrees. Extrapolation of the particle diameter from the image reveals that they are in contact at the equator, shown in the image as white circles. The white solid in the image is crystalline water vapour condensed from the atmosphere during sample loading into the SEM. Scale bar 3 µm.

Samples were viewed and imaged in a Phillips Leo Scanning Electron Microscope (SEM).

**Exposure of non-spherical bubbles to surfactants:** 50 µl of Triton-X100 (Sigma) solution at a concentration of 1.7 mM was introduced into the glass chamber and allowed to diffuse to the armored bubbles. Movies of the destabilization process were taken with a Phantom V5 high-speed camera (Vision Research).

## Results and Discussion

**Interfacial Composite Materials:** The solid-like properties of armored interfaces manifest themselves when the underlying fluid/fluid interface is constrained to an area larger than its equilibrium value, due to the presence of the adsorbed particles. Indeed, in this regime, the area minimizing fluid/fluid interface exerts a capillary force on the particles, which causes the steric jamming of the particles[6, 20]. This fundamental property of jamming on fluid interfaces seems to be independent of the particle and fluid type[3].

As we recently reported, particulate interfaces with excess area can be created through the fusion of two spherical bubbles armored with smaller spherical particles[3] (Fig. 2a). Unlike ordinary fluid bubbles which remain spherical and decrease their surface area by about 20 percent after fusion (for bubbles of the same initial size), the particles constrain the area of the interface and thus the resultant bubble maintain an ellipsoidal shape. This conservation of both surface area and volume in particle-covered bubbles holds true even when there are many parent bubbles or even when the bubbles are non-identical. Thus extremely high-aspect ratio objects, such as the spherocylindrical bubbles shown in Figure 2b can be created by fusion of many spherical armored bubbles. The ICM is clearly behaving in a solid-like manner, since the long cylinder does not undergo a Rayleigh-Plateau capillary instability.

Alternatively, for bubbles with only a small number of adsorbed particles, the jamming threshold can also be reached by reducing the volume of a single armored bubble. An initial low surface density of particles allows thermal motion of the particles along the interface. When these bubbles are reduced in volume, as occurs naturally owing to dissolution of the gas in liquid, Brownian motion of the particles is arrested (i.e. the particles jam) as the surface area of the bubble decreases. Further volume reduction leads to crumpled structures such as the membrane-like bags shown in Fig. 2c, which were obtained by partially evacuating bubbles armed with 200 nm particles. Similar buckled structures were obtained by Xu et. al by removing some of the oil in particle-covered oil droplets[5], and were previously noticed in other settings[7, 10]. The buckling response is thus also a generic property of ICMs[5, 16].

Once the interface is jammed, the bubble can not only support high-aspect ratio and crumpled shapes but is also able to support both concave and convex curvatures on the same bubble and a change in bubble topology (Fig. 2d)[3], all of which are prohibited for an ordinary bubble. To better understand these observations, we next consider the microstructure and mechanical response of the ICM.

**Microstructure of the ICM and its mechanical response to stresses:** Our initial attempts to obtain high-resolution SEM images of the ICM microstructure through standard methods failed because we found that the drying of an aqueous suspension of the armored bubbles leaves a cohesionless residue of particles. This observation is noteworthy since it reveals the lack of strong bonds between the particles, which is an important distinguishing feature between the ICM and more familiar atomic solids. Moreover, it is clear that any solid-like characteristic of the ICM is mediated by the presence of the liquid-gas interface.

Thus, to ascertain the structure of the ICM in its natural state, we freeze-fractured a solution of armored bubbles and visualized the bubbles using a cryo-SEM. The rapid freezing of the bubbles preserves the structure of the fluid/fluid interface and the adsorbed particles (Fig. 3a,b). It is clear that the particles are adsorbed on the interface with a well-defined contact line, and the particles are hexagonally close-packed with only a few defects (Fig. 3a,b). Moreover, we found that the particles have a rather low contact angle of 40 degrees, which demonstrates that interfacial jamming is possible with non-neutrally wetting particles.

Our microstructural observations indicate that the ICM is a two-dimensional granular solid. Thus, it is to be expected that the ICM should have properties in common with other granular systems. In more familiar jammed two-dimensional planar systems, and for three-dimensional systems more generally, force chains have been proposed to serve as the load-bearing network that spans the boundaries of the system[21, 22], and give rise to solid-like resistance to shear. However, jammed systems in open geometries are extremely 'fragile', since the force chains are easily destroyed by incompatible stresses at the boundaries[21]. In the case of armored bubbles, equilibrium demands that force chains must be closed (for the bubbles are closed compact surfaces). Thus, the armored bubbles should be unusually resistant when subject to homogenous stresses. A simple verification of this idea is afforded by subjecting the bubbles to simple shear flow. The bubbles tumbled like a solid object, and did not transition to tank-treading[23] (a liquid-like response) with higher shear rates, reflecting its significant shear resistance (Supplementary Movie 1). Because of the rigid body rotation, every surface element of the ICM experiences the same average stress.

We next probe the mechanical response of the ICM to inhomogeneous stresses by submitting a prolate ellipsoidal armored bubble to uniaxial compression along its major axis in a glass chamber (Fig. 4a). Compression causes the bubble to transition to a cuboidal shape before eventually assuming an oblate shape (Fig. 4b,c,d). When the glass plates are removed from contact with the bubble, the oblate ellipsoidal bubble is stable. Direct observations of the particles on the interface show that they slide in 45 degree shear bands relative to the direction of compression, which, as we discuss below, reveals stress localization in the ICM.

The prolate to oblate change in equilibrium shape we show here, which is caused by an external load, is termed plastic deformation in solid mechanics. It is thus clear that the bubble undergoes extensive plastic deformation when subject to uniaxial compression. We note that no particles are observed to leave the colloidal monolayer during this process, and the plastic deformation that we show here (i.e. gross changes in shape, from one static shape to another static shape) can be performed repeatedly with no apparent hindrance to the process. These results demonstrate that the final configuration of the ICM, and thus the bubble shape, is only dependent on immediate strain history and is independent of past strain history. In contrast, atomic solids usually undergo strain hardening or other irreversible changes when plastically strained, which leads to solid fatigue, and a permanent change in mechanical response. The apparent fatigue-free deformation of ICMs may be a general response, since similar behavior is observed for the compression and dilation of particle-covered droplets[5] and planar monolayers[16].

We discuss next how this experimental verification of the rigidity of the armored bubbles under nearly homogenous stress as opposed to the apparent softness of the material under localized compression is linked directly to the microstructure of the ICM, and its ability to manifest solid-like properties in spite of the absence of permanent bonds between the particles. In isotropic compression, the theoretical response of the particulate monolayer is determined by the mechanics of contact between the particles, while in isotropic tension the stresses are borne by the cost of creating excess air/water interface, both of which are reversible and elastic in nature, i.e. the material returns to its original configuration when the external stress is released. Thus, the plasticity of the armored bubble must arise from shear deformations. For example, in the inset of Fig. 4e we show a unit cell of three close-packed disks of radius $r$ confined to a plane. It is apparent that a pure shear must cause an area dilatation of the unit cell to allow the relative translation of the close-packed particles, which is analogous to the volume dilatation of granular media in three dimensions[24]. Elementary geometry shows that the microscopic yield strain associated with this dilatation of the solid is $\varepsilon_{micro}=(2-\sqrt{3})/\sqrt{3} \sim 0.15$, i.e. this is the strain required to allow plastic flow (Fig. 4e, inset). This microscopic yield strain is rather large since, for example, most atomic solids fail when their microscopic strains exceed 0.02[31]. Once this yield strain is reached, the particles would flow on the interface so that the material strains continuously at a constant applied stress. This liquid-like response is termed perfectly plastic deformation.

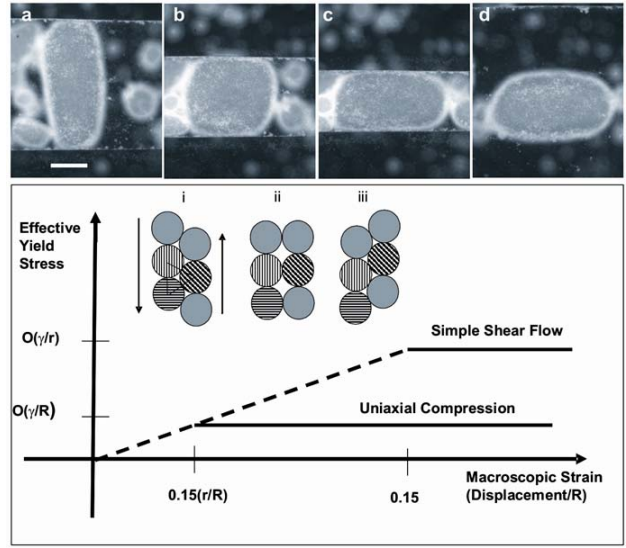

**Figure 4:** Mechanical response of the ICM. (a) A prolate ellipsoidal bubble subjected to compression in a glass chamber. (b,c) The ICM undergoes plastic rearrangement to accommodate the uniaxial compressive stress. This plastic accommodation is extensive, and the bubble compresses to a cuboidal shape before extending a major axis orthogonal to that of the pre-stressed object. (d) When the side plates are removed from direct contact with the bubble, the ellipsoid produced is stable in its new configuration. Ejection of individual particles from the ICM is not observed during any of the above manipulations. Scale bar 200 μm. (e) **inset, i,** Shearing of a hexagonally close-packed (HCP) monolayer of rigid particles is impossible at constant surface area. Shear stress results in the dilation of the surface. **ii,** The relative displacement, or strain is $(2-\sqrt{3})/\sqrt{3} = 0.15$ when the particles are square-packed, which is the microscopic strain. **iii,** A little past this point, the work put in to create the excess surface area is regained elastically as the particles assume HCP once again. Thus, the area strain at the end of the deformation is zero, but the unit cell configuration has changed. For a bubble subject to a uniform applied stress, such as in simple shear flow, a solid-to-liquid transition requires a microscopic strain over most of the bubble surface, which correspondingly requires a large yield stress about $O(\sigma/r)$, where $\sigma$ is the air/water surface tension and $r$ is the particle radius. In contrast, when the applied stress is non-uniform over the surface of a bubble of radius $R$, such as in uniaxial compression, stresses can be focused along shear bands, and now the effective yield stress required for plastic deformation is much lower, $O(\sigma/R)$, since the apparent yield strain distributed over the bubble surface is 0.15(r/R).

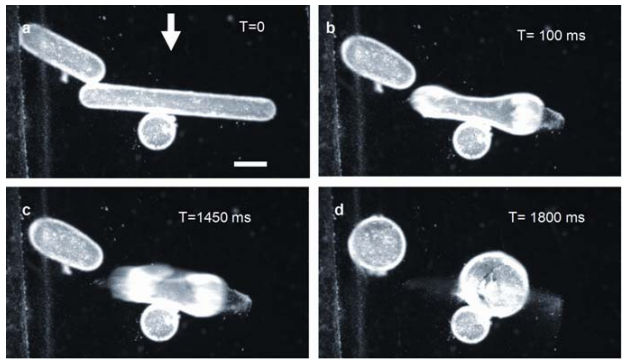

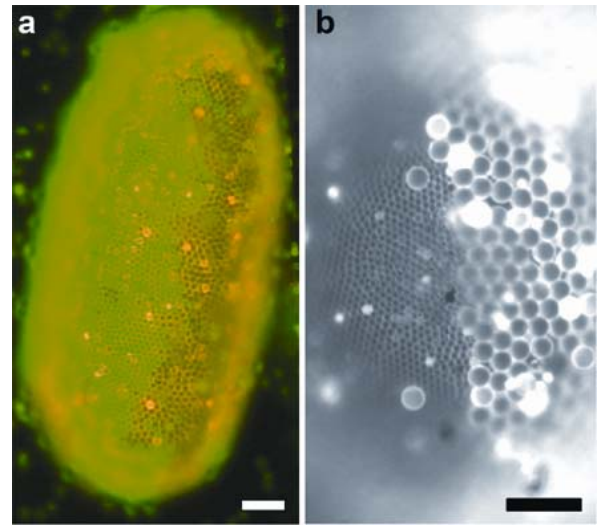

**Figure 5:** Stress distribution in a spherocylindrical ICM bubble. The bubbles are coated with 2.6 μm diameter charge-stabilized fluorescent polystyrene beads. The surfactant employed was Triton-X-100 at a concentration of 1.7 mM. (a) The initial state, with the white arrow indicating the direction of surfactant arrival. For an elastic spherocylinder, the hoop stress is twice as large as the axial stress. (b, c) Failure begins at the axial spherical caps, where the ejection of particles is first observed. An opening of the cylindrical region relieves the initially anisotropic stress: decreasing the radius of curvature in the hoop direction and increasing the radius of curvature in the axial direction, until (d) they are both equal, i.e. when the bubble becomes a sphere. Scale bar 300 μm.

**Figure 6:** The composite nature of the ICM is not limited to the properties of the fluid interface and the particles, in (a) a Janus ellipsoid is created from two kinds of fluorescently-labelled polystyrene beads. Scale bar 40 μm. (b) A close up of a Janus ellipsoid with two different sizes of particles on the interface. Since the mechanical properties of the ICM is determined by the particle size, varying the particle size may affect the properties and stress distribution on this object. Scale bar 20 μm.

It now follows that the macroscopic strain for this system is $\varepsilon_{macro} = \varepsilon_{micro} \, r/R \ll \varepsilon_{micro}$ where $R$ is the radius of the armored bubble. This suggests that if the stresses are inhomogeneous on the armor, which occurs during uniaxial compression of the bubbles between glass plates (Fig. 4a-d), the microscopic stresses can now be large enough to cause localized plastic deformation or yielding along shear bands, i.e. there are large local strains. In contrast, with an (approximately) homogenous applied stress, such as when the armored object tumbles in a uniform shear flow, plastic deformation globally would require a macroscopic yield strain on the order of $\varepsilon_{micro}$ over most of the surface (Fig. 4e) (Supplementary Information). Thus, we expect that localized events dominate the microscopic response to large-scale forces. The large microscopic yield strain also prevents spontaneous rearrangement of the particles on the ICM. Indeed, the normal forces that arise due to the jamming negate the need for friction in maintaining the non-spherical shapes in static equilibrium. We note that although friction is not necessary to maintain the jammed state, it may be important in the dynamic process of reconfiguration when particles slide past one another in response to an external stress.

The above characterization makes it clear that the jamming of particles on the interface supports the many stable non-spherical and/or folded shapes that we have observed; see also[4]. This description is independent of the detailed attractive or repulsive potentials of the particles and is qualitatively consistent with our experiments since we observe similar behavior for a wide variety of particle types. It is interesting to note that bulk rheological measurements of flocculated particle-stabilized emulsions[25], and surface rheological measurement of interfaces covered with a high concentration of large molecular weight proteins also exhibit mechanical responses consistent with jamming[26].

**Determination of stress distribution on the ICM:** Although, the shape of the non-spherical bubble implies anisotropic stresses on the ICM, it is still unclear how the stresses are distributed. Indeed, there is still much debate on how stresses in granular media are supported and transmitted[27-29]. Here, the non-trivial geometry of a non-spherical bubble can be exploited to obtain insights into the stress state of the particulate monolayer. Accordingly, we created high-aspect ratio spherocylindrical bubbles and destabilized them by exposure to a non-ionic surfactant, which causes the ejection of particles from the interface[30].

We observe that the spherocylindrical bubble destabilizes at the axial ends and grows spherical caps, which eventually collapses to form a sphere (Fig. 5). In a thin elastic spherocylinder, the cylindrical region has a hoop stress resultant, $\sigma_h = \dfrac{R\Delta P}{2r}$ that is twice as large as the axial stress resultant, $\sigma_a = \dfrac{R\Delta P}{4r}$, where $\Delta P$ is the pressure difference between the bubble and the surroundings[31], while the hemispherical end caps have a stress resultant of $\dfrac{R\Delta P}{2r}$. This would suggest that at the hemisphere-cylinder junction, there is a stress discontinuity that causes a local barreling effect due to the jump in curvature. Indeed, in the experiments, the cylinder fails at or near the stress discontinuity (Fig. 5a-d), and the anisotropy in stress on the interface is relaxed by simultaneously decreasing the hoop radius of curvature and increasing the axial radius of curvature, and stops when they become equal, i.e. when the bubble becomes a sphere.

### Conclusion

We have provided experimental and theoretical characterizations of the recently reported interfacial jamming phenomena, which leads to solid-like properties in particle-covered interfaces. The solid-like characteristics arise solely due to steric jamming, and the jammed material can rearrange plastically when subject to inhomogeneous external stresses, and can take on various stable configurations. A unique property of jamming on closed interfaces is the apparent resistance of the ICM to homogenous stresses. We also show that in simple geometries, the stress distribution on the ICM is consistent with that of a conventional thin elasto-plastic shell.

An implication of the 'composite nature' of the ICM is that additional degrees of freedom can be obtained by varying the size and type of particles used prior to the fusion process. Hence we have been able to make Janus ellipsoids (Fig. 6). Since the mechanical properties of the ICM is directly related to the particle size, having sections with varying particle sizes should give rise to further novel mechanical behavior. Finally, the successful production and stable maintenance of non-spherical bubbles opens experimental avenues for these systems as model two-dimensional crystalline surfaces[32, 33], and offers the potential for new ideas for bubble transport and delivery[34, 35].

## Acknowledgements.
We thank the Harvard Material Research Science and Engineering Center (DMR-0213805) and Unilever Research for support. We also thank P. Dimotakis for the observations of long-lived gas bubbles in the ocean, R. Schalek for help with the SEM imaging, and D. Gregory and colleagues for support and helpful conversations.

**Supporting Information Available:** Movie showing the rigid body tumbling of an armored bubble in shear flow. Additional supporting calculations. This material is available free of charge via the Internet at http://pubs.acs.org.